\documentclass[letterpaper, 12pt]{article}

\usepackage[hmargin=3.17cm,vmargin=2.54cm]{geometry}

\usepackage{graphicx}
\usepackage{amsfonts}
\usepackage{amssymb}
\usepackage{amsmath}
\usepackage{booktabs}
\usepackage{doi}
\usepackage{setspace}
\begin{document}

\title{Inverse design and demonstration of a compact and broadband on-chip wavelength demultiplexer}
\author{Alexander Y. Piggott$^1$, Jesse Lu$^1$, Konstantinos G. Lagoudakis$^1$, \\ Jan Petykiewicz$^1$, Thomas M. Babinec$^1$, and Jelena Vu\v{c}kovi\'{c}$^{1\ast}$}
\date{}
\maketitle

\begin{center}
\vspace{-4ex}
$^1$ Ginzton Laboratory, Stanford University, Stanford, CA, 94305

$^\ast$ \emph{jela@stanford.edu}
\end{center}

\begin{abstract}
Integrated photonic devices are poised to play a key role in a wide variety of applications, ranging from optical interconnects \cite{dabmiller_ao2010} and sensors \cite{vsylin_sci1997} to quantum computing \cite{pkok_rmp2007}. However, only a small library of semi-analytically designed devices are currently known \cite{gtreed_2008}. In this paper, we demonstrate the use of an inverse design method that explores the full design space of fabricable devices and allows us to design devices with previously unattainable functionality, higher performance and robustness, and smaller footprints compared to conventional devices \cite{jlu_oe2013}. We designed a silicon wavelength demultiplexer that splits $1300~\mathrm{nm}$ and $1550~\mathrm{nm}$ light from an input waveguide into two output waveguides, and fabricated and characterized several devices. The devices display low insertion loss $\left(2 - 4~\mathrm{dB}\right)$, high contrast $\left(12 - 17~\mathrm{dB}\right)$, and wide bandwidths $\left(\sim 100~\mathrm{nm} \right)$.  The device footprint is $2.8 \times 2.8 ~\mathrm{\mu m}$, making this the smallest dielectric wavelength splitter to date.
\end{abstract}

Electronic hardware description languages, such as Verilog and VHDL, are widely used in industry to design digital and analog circuits \cite{pjashenden_2008, ieeestd_verilog}. The automation of large scale circuit design has enabled the development of modern integrated circuits which can contain billions of transistors. Photonic devices, however, are effectively designed by hand. The designer selects an overall structure based on analytic theory and intuition, and then fine tunes the structure using brute-force parameter sweep simulations. Due to the undirected nature of this process, only a few degrees of freedom $\left(2 - 6\right)$ are available to the designer.  The field of integrated photonics would be revolutionized if the design of optical devices could be automated to the same extent as circuit design.

We have previously developed an algorithm which can automatically design arbitrary linear optical devices \cite{jlu_oe2013}. Our method allows the user to \emph{design by specification}, whereby the user simply specifies the desired functionality of the device, and the algorithm finds a structure which meets these requirements. In particular, our algorithm searches the full design space of fabricable devices with arbitrary topologies. These complex, aperiodic structures can provide previously unattainable functionality, or higher performance and smaller footprints than traditional devices, due to the greatly expanded design space \cite{jlu_oe2013, jsjensen_apl2004, piborel_oe2004, amutapcica_eo2009, jjensen_lpr2011, lalau-keraly_oe2013, aniederberger_oe2014, aypiggott_sr2014}. Our algorithm uses local-optimization techniques based on convex optimization \cite{sboyd_2004} to efficiently search this enormous parameter space. 

Here, we demonstrate the capabilities of our inverse design algorithm by designing and experimentally demonstrating a compact wavelength demultiplexer on a silicon-on-insulator (SOI) platform. One of the key functions of silicon photonics is wavelength division multiplexing (WDM), which multiplies the data capacity of a single optical waveguide or fiber optic cable by the number of wavelength channels used \cite{fxia_oe2007, qfang_oe2010, aalduino_iprsn2010}. Unfortunately, conventional wavelength demultiplexers such as arrayed waveguide gratings \cite{ksasaki_el2005}, echelle grating demultiplexers \cite{fhorst_iptl2009}, and ring resonator arrays \cite{msdahlem_oe2011} are fairly large, with dimensions ranging from tens to hundreds of microns \cite{wbogaerts_ijqe2010}. Our device has a footprint of only $2.8 \times 2.8 ~\mathrm{\mu m}$, which is considerably smaller than any previously demonstrated dielectric wavelength splitter \cite{lhfrandsen_cleo2013}.

Let us now consider the general formulation of the inverse design problem for optical devices. We choose to specify performance of our device by defining the mode conversion efficiency between sets of input modes and output modes at several discrete frequencies. These modes and frequencies are specified by the user, and kept fixed during the optimization process. In the limit of a continuous spectrum of frequencies, any linear optical device can be described by the coupling between sets of input and output modes, making this a remarkably general formulation \cite{dabmiller_oe2012}.

Suppose the input modes $i = 1 \ldots  M$ are at frequencies $\omega_i$, and can be represented by equivalent current density distributions $\mathbf{J}_i$. Then the electric fields $\mathbf{E}_i$ generated by the input modes should satisfy Maxwell's equations in the frequency domain,
\begin{align}
\nabla \times \mu_0^{-1} \nabla \times \mathbf{E}_i - \omega_i^2 \, \epsilon \, \mathbf{E}_i = - i \omega_i \mathbf{J}_i, \label{eqn:wgWDM_physics_spec_1}
\end{align}
where $\epsilon$ is the electric permittivity, and $\mu_0$ is the magnetic permeability of free space.

We can then specify $N_i$ output modes of interest for each input mode $i$. We define the output mode electric fields $\mathcal{E}_{ij}$ over output surfaces $S_{ij}$, where $j = 1 \ldots N_i$. The device performance is then specified by constraining the amplitude coupled into each output mode to be between  $\alpha_{ij}$ and $\beta_{ij}$. This leads to the constraint,
\begin{align}
\alpha_{ij} \leq \left| \iint_{S_{ij}} \mathcal{E}^\dagger_{ij} \cdot \mathbf{E}_i \mathrm{d}S \right| \leq \beta_{ij} \label{eqn:wgWDM_physics_spec_2}
\end{align}
where we have used overlap integrals to compute the mode coupling efficiency into each output mode, and assumed that the input and output modes are appropriately normalized.

The inverse design problem thus reduces to finding the permittivity $\epsilon$ and electric fields $\mathbf{E}_i$ which simultaneously satisfy physics, described by equation \ref{eqn:wgWDM_physics_spec_1}, and the device performance constraints, described by equation \ref{eqn:wgWDM_physics_spec_2}. In general, we also have additional constraints on the permittivity $\epsilon$ due to fabrication limitations.

%Naive approaches to solving this problem, such as genetic algorithms \cite{agondarenko_oe2008, ahakansson_oe2005,mminkov_sr2014} and particle swarm optimization \cite{yma_oe2013}, ignore the underlying physics and scale poorly with additional degrees of freedom for the structure. These methods require approximately one electromagnetic simulation for each additional degree of freedom on every iteration. More sophisticated approaches, such as those used by our inverse design algorithm, take advantage of the underlying physics \cite{jsjensen_apl2004, piborel_oe2004, amutapcica_eo2009, jjensen_lpr2011, lalau-keraly_oe2013, aniederberger_oe2014}. These methods generally require on the order of one electromagnetic simulation for each input and output mode every iteration; the computational cost is essentially independent of the number of degrees of freedom.

We use two methods for solving this problem, the \emph{objective first} method \cite{jlu_oe2013} and a \emph{steepest descent} method. In the \emph{objective first} method, we constrain the electric fields $\mathbf{E}_i$ to satisfy our performance constraints in equation \ref{eqn:wgWDM_physics_spec_2}, but allow Maxwell's equations to be violated. We then minimize the violation of physics using the Alternating Directions Method of Multipliers (ADMM) optimization algorithm \cite{jlu_oe2013}. We call this method ``objective first'' since we are forcing the fields to satisfy the performance objectives first, and then attempting to satisfy Maxwell's equations.

In our \emph{steepest descent} method, we constrain our electric fields $\mathbf{E}_i$ to satisfy Maxwell's equations, and define a performance metric function based on the violation of our device performance constraints in equation \ref{eqn:wgWDM_physics_spec_2}. We then compute the local gradient of the performance metric by solving an adjoint electromagnetic problem, and perform steepest-gradient descent optimization \cite{jlu_oe2013,aypiggott_sr2014}.

\begin{figure}
	\center
	\includegraphics[scale=0.45]{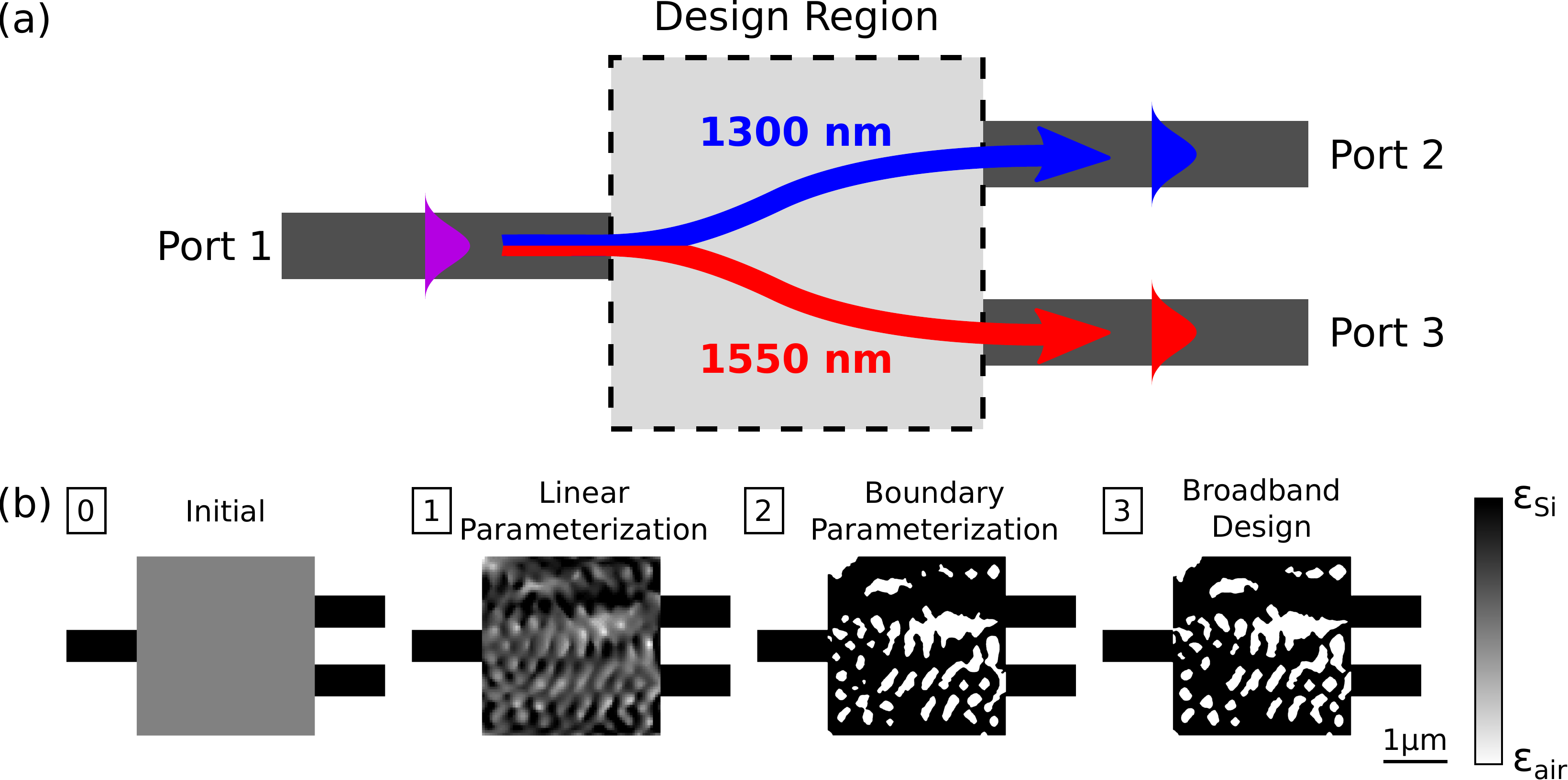}
	\caption{Overview of the inverse design process. \textbf{(a)} The device functionality is defined for the inverse design algorithm by specifying the surrounding structure, the design region, and the coupling between a set of input and output modes. For the compact wavelength demultiplexer demonstrated in this work, the structure consists of one input waveguide, two output waveguides, and a $2.8 \times 2.8 ~\mathrm{\mu m}$ design region. $1300~\mathrm{nm}$ band light is coupled into the fundamental TE mode of port 2, and $1550~\mathrm{nm}$ band light is coupled into the fundamental TE mode of port 3. All three waveguides are identical, with a width of $500~\mathrm{nm}$. \textbf{(b)} Intermediate structures generated by the inverse design process. In the first stage, the structure is optimized while allowing the permittivity $\epsilon$ to continuously vary (linear parameterization). In the next stage, we convert to a boundary parameterization and optimize the structure for operation at only $1300~\mathrm{nm}$ and $1550~\mathrm{nm}$. In the final stage, we perform a broadband optimization to generate a robust device. \label{fig:1_wgWDM_invdes}}
\end{figure}

To design the compact wavelength demultiplexer, we chose a simple planar 3-port structure with one input waveguide, two output waveguides, and a square design region, as illustrated in figure \textbf{\ref{fig:1_wgWDM_invdes}a}. For ease of fabrication, the structure was constrained to a single fully etched $220~\mathrm{nm}$ thick $\mathrm{Si}$ layer on a $\mathrm{Si}\mathrm{O}_2$ substrate with air cladding. Refractive indices of $n_\mathrm{Si} = 3.49$, $n_{\mathrm{Si}\mathrm{O}_2} = 1.45$, and $n_\mathrm{Air} = 1$ were used. The fundamental TE mode of the input waveguide was used as the input mode for the inverse design procedure, and the fundamental TE modes of the two output waveguides were used as the output modes. At $1300~\mathrm{nm}$, we specified that $>90\%$ of the input power should be coupled out of port 2 and $<1\%$ of the power should be coupled out of port 3; the converse was specified for $1550~\mathrm{nm}$.

The optimization processes proceeded in several stages, as outlined in figure \textbf{\ref{fig:1_wgWDM_invdes}b}. In the first stage, the permittivity $\epsilon$ in the design region was allowed to vary continuously between the permittivity of silicon and air (linear parameterization). The objective first method was used to generate an initial guess for the structure, and then the steepest descent method was to fine tune the structure. In the second stage, the structure was converted to a binary level-set representation \cite{sosher_2003}, and then optimized using steepest descent. Up to this point, the device performance had only been specified at the two center wavelengths, $1300~\mathrm{nm}$ and $1550~\mathrm{nm}$. In the final stage, the device was optimized for broadband performance by specifying the device performance at 10 different wavelengths, with 5 frequencies equally spaced about each center frequency. Broadband performance was previously shown to be a heuristic for structures which are tolerant to fabrication imperfections, and it was hoped that this would result in a more robust design \cite{jlu_oe2013}. The WDM device was designed in approximately 36 hours using a single server with three NVidia GTX Titan graphics cards. 

\begin{figure}
	\center
	\includegraphics[scale=0.50]{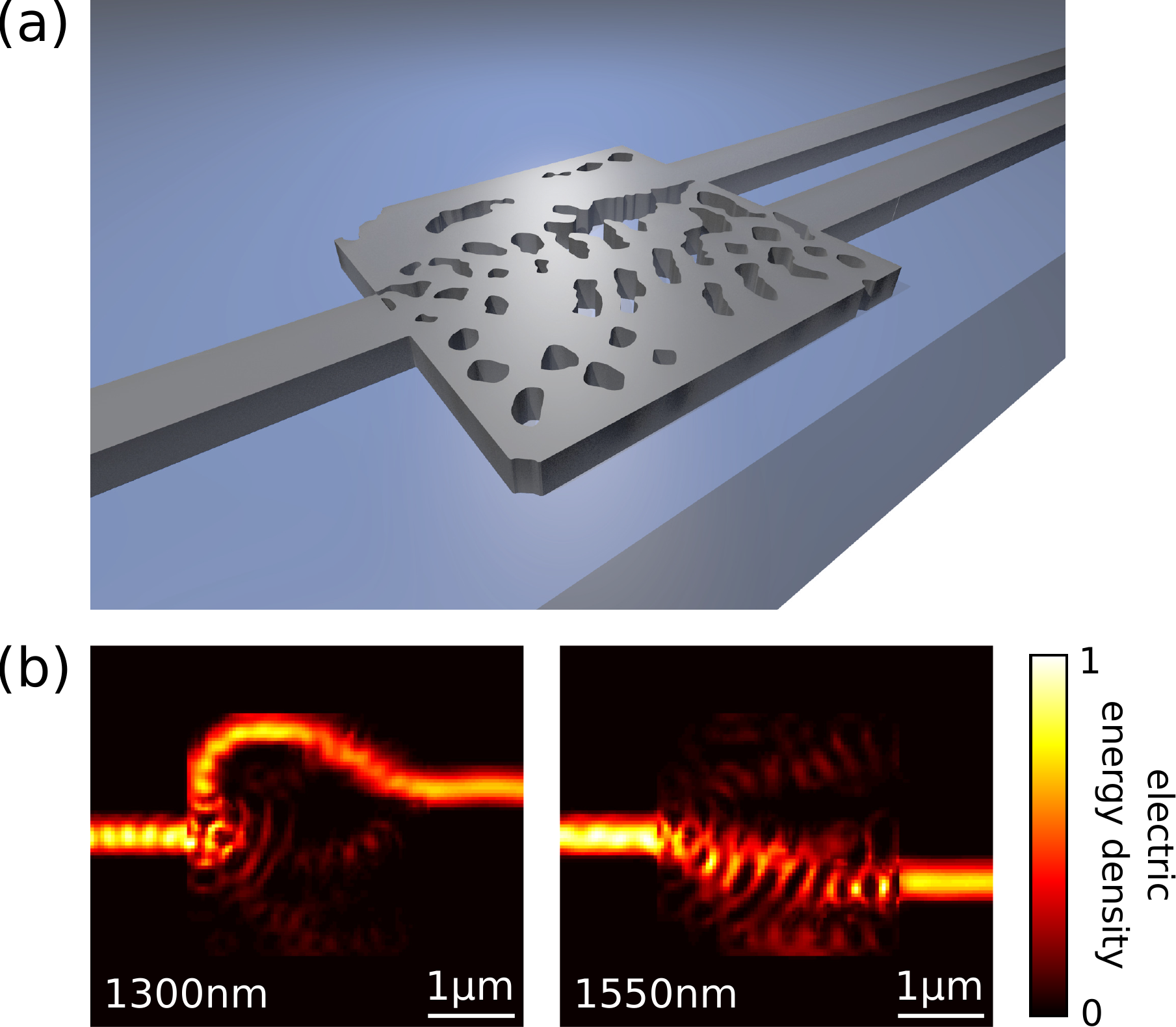}
	\caption{The compact wavelength demultiplexer designed by the inverse design algorithm. \textbf{(a)} A three-dimensional rendering of the structure. Silicon is coloured grey, and $\mathrm{Si}\mathrm{O}_2$ is coloured blue. \textbf{(b)} Field plots of the device operating at $1300~\mathrm{nm}$ and $1550~\mathrm{nm}$. Here, we have plotted the electric energy density $U = \epsilon | \vec{E} |^2$.\label{fig:2_wgWDM_struct}}
\end{figure}

The final designed device is shown in \textbf{\ref{fig:2_wgWDM_struct}a}. The simulated electric fields at the central operating wavelengths of $1300~\mathrm{nm}$ and $1550~\mathrm{nm}$ are plotted in figure \textbf{\ref{fig:2_wgWDM_struct}b}. At both wavelengths, the light takes a relatively confined path through the structure, despite the convoluted geometry of the etched silicon layer.

\begin{figure}
	\center
	\includegraphics[scale=0.50]{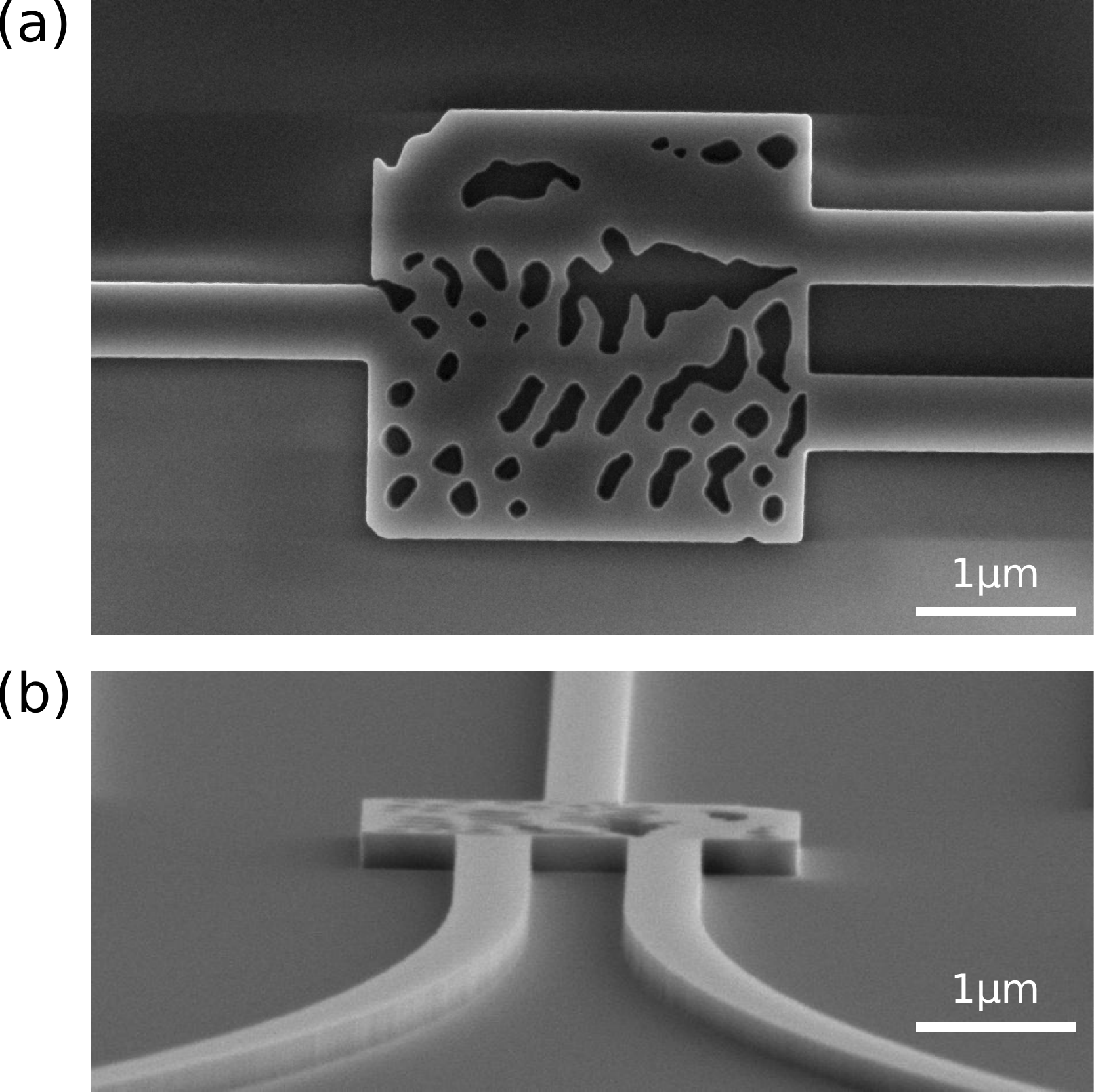}
	\caption{Scanning electron microscopy (SEM) images of the fabricated wavelength demultiplexer. The device was fabricated by fully etching the $220~\mathrm{nm}$ thick device layer of a silicon-on-insulator (SOI) substrate. \textbf{(a)} Top down view. \textbf{(b)} Angled view. The vertical sidewalls are clearly visible in this view. \label{fig:3_wgWDM_SEM}}
\end{figure}

The devices were fabricated by using electron beam lithography followed by plasma etching. Scanning electron microscopy (SEM) images of a final fabricated device is shown in figure \textbf{\ref{fig:3_wgWDM_SEM}a}. The original design was accurately reproduced by the fabrication process, with the exception of two small $\left(\sim 100~\mathrm{nm}\right)$ holes next to the input waveguide which are missing.

\begin{figure}
	\center
	\includegraphics[scale=0.50]{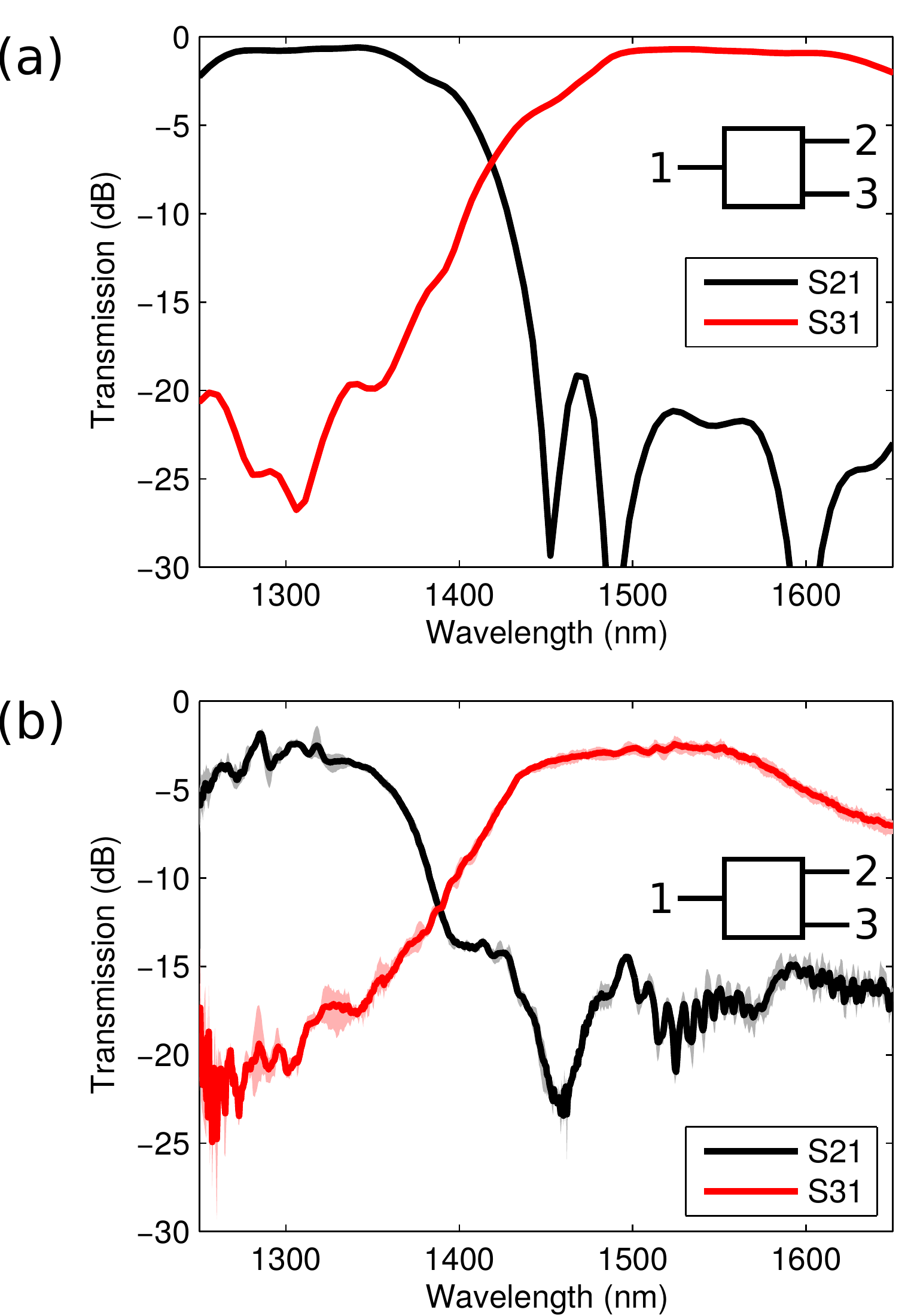}
	\caption{S-parameters for the device. Here, we have plotted transmission from input port 1 to output ports 2 and 3. \textbf{(a)} S-parameters simulated using finite-difference frequency-domain (FDFD) simulations. \textbf{(b)} Measured S-parameters for 3 identical devices. The shaded areas indicate the minimum and maximum measured values across all measured devices, and solid lines indicate the average values. The insertion losses and contrast of the device are somewhat degraded with respect to the simulated values due to fabrication imperfections. \label{fig:4_wgWDM_S-param}
}
\end{figure}

The measured and simulated S-parameters for the compact WDM device are plotted in figure \textbf{\ref{fig:4_wgWDM_S-param}}. The plotted wavelength range was limited by the spectral bandwidth of the excitation source. Measurements from three identically fabricated devices are plotted together in figure \textbf{\ref{fig:4_wgWDM_S-param}b}, showing that device performance is highly repeatable.  Although somewhat degraded in performance with respect to the simulated devices, the fabricated WDM devices exhibit relatively low insertion loss $\left(2 - 4~\mathrm{dB}\right)$, high contrast $\left(12 - 17~\mathrm{dB}\right)$, and very broadband $\left(\sim 100~\mathrm{nm} \right)$ pass and stop bands. We attribute the discrepancy between simulation and measurement to fabrication imperfections. 

In summary, we have experimentally demonstrated a compact, practical wavelength demultiplexer designed using our inverse design algorithm. This device provides functionality which never before been demonstrated in such a small structure. Our results suggest that the inverse design of optical devices will revolutionize integrated photonics, ushering in a new generation of highly compact optical devices with novel functionality and high efficiencies.

\section*{Methods}

\subsection*{Optimization Algorithm and Electromagnetic Simulations}
The detailed inverse design algorithm has been previously described elsewhere \cite{jlu_oe2013,aypiggott_sr2014,jlu_thesis2013}. A graphical processing unit (GPU) accelerated implementation of the MaxwellFDFD finite-difference frequency-domain solver was used to efficiently solve Maxwell's equations throughout the optimization process \cite{wshin_jcp2012, wshin_maxwell_webpage}.

\subsection*{Fabrication}
The devices were fabricated using Unibond$^{\mathrm{TM}}$ SmartCut$^{\mathrm{TM}}$ silicon-on-insulator (SOI) wafers obtained from SOITEC, with a nominal $220~\mathrm{nm}$ device layer and $3.0~\mathrm{\mu m}$ BOX layer. A JEOL JBX-6300FS electron beam lithography system was used to pattern a $330~\mathrm{nm}$ ZEP-520A electron beam resist layer spun on the samples. We did not compensate for the proximity effect in the electron beam lithography step. A transformer-coupled plasma (TCP) etcher was used to transfer the mask to the silicon device layer, using a $\mathrm{C}_2 \mathrm{F}_6$ breakthrough step and a $\mathrm{B}\mathrm{Cl}_3 / \mathrm{Cl}_2  / \mathrm{O}_2$ chemistry main etch. The mask was stripped by soaking in Microposit Remover 1165, followed by a piranha clean using a $4:1$ ratio of concentrated sulfuric acid and $30\%$ hydrogen peroxide. Finally, the samples were diced and polished to expose the waveguide facets for edge coupling. Detailed schematics of the device are available in the supplementary information.

\subsection*{Measurement}
The devices were measured by edge-coupling the input and output waveguides to lensed fibers. A polarization maintaining (PM) lensed fiber was used on the input side to ensure that only the fundamental TE waveguide mode was excited. The polarization extinction ratio of the light emitted by the PM lensed fiber was measured using a polarizing beamsplitter to be $19.0~\mathrm{dB}$ at $1470~\mathrm{nm}$, and $20.7~\mathrm{dB}$ at $1570~\mathrm{nm}$. A non-polarization maintaining lensed fiber was used to collect light from the outputs.  The lensed fibers were aligned by optimizing the transmission of a laser at $1470~\mathrm{nm}$, ensuring consistent coupling regardless of the transmission characteristics of the devices. 

A fiber-coupled broadband light-emitting diode (LED) source and fiber-coupled optical spectrum analyzer (OSA) were used to characterize the devices. The transmission measured through each device was normalized with respect to a straight-through waveguide running parallel to each device. This eliminated any coupling and waveguide losses, and yielded a direct measurement of the device efficiencies.

\section*{Acknowledgements}
This work was supported by the AFOSR MURI for Complex and Robust On-chip Nanophotonics (Dr. Gernot Pomrenke), grant number FA9550-09-1-0704. A.Y.P. also acknowledges support from the Stanford Graduate Fellowship. K.G.L. acknowledges support from the Swiss National Science Foundation. J.P. was supported in part by the National Physical Science Consortium Fellowship and by stipend support from the National Security Agency.

We would like to acknowledge Prof. Stephen Boyd for his theoretical guidance and fruitful discussions regarding the optimization algorithm. In addition, we would like to thank Prof. Joyce Poon for her generous donation of the SOI wafer used to fabricate our devices.

\section*{Author contributions}
A.Y.P. designed, simulated, fabricated, and measured the devices. J.L. developed the inverse design algorithm and software. K.G.L. assisted building the measurement setup, J.P. contributed to the simulation software, and T.B. provided theoretical and experimental guidance. J.V. supervised the project. All members contributed to the discussion and analysis of the results.

\section*{Competing financial interests}
The authors declare no competing financial interests.

\clearpage

\clearpage
\begin{center}
\LARGE{Supplementary Information}
\end{center}

\section{Device Schematic}
\begin{figure}[h]
	\center
	\includegraphics[width=0.8\textwidth]{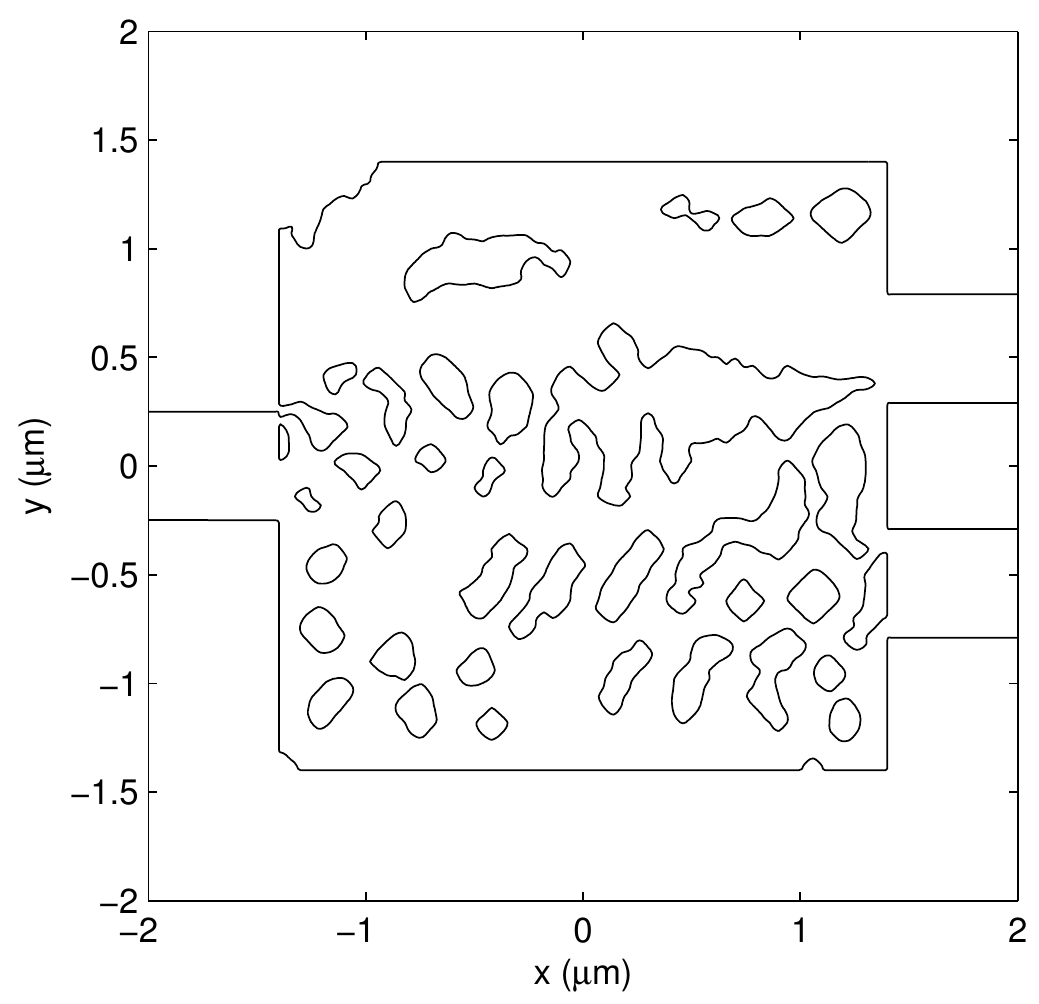}
	\caption{Detailed schematic of the compact wavelength demultiplexer.}
\end{figure}

\end{document}